\documentclass[aps,prl,twocolumn,showpacs,floatfix,superscriptaddress]{revtex4-1}

\usepackage{times,mathrsfs}
\usepackage{graphicx,color}
\usepackage{amsmath,amssymb}
\usepackage[colorlinks,citecolor=blue]{hyperref}

\begin{document}

\newcommand\hatH{{\hat{H}}}
\newcommand\calQ{{\mathcal{Q}}}

\newcommand\jj{\color{blue}}

\title{
{Thermodynamic uncertainty relation of interacting oscillators in synchrony}}

\author{Sangwon Lee}
\affiliation{Department of Physics and Astronomy, Seoul National University, Seoul, Korea}

\author{Changbong Hyeon}
\affiliation{Korea Institute for Advanced Study, Seoul, Korea}

\author{Junghyo Jo} \email{jojunghyo@kias.re.kr}
\affiliation{Korea Institute for Advanced Study, Seoul, Korea}

\date{\today}

\begin{abstract}
The thermodynamic uncertainty relation sets the minimal bound of the cost-precision trade-off relation for dissipative processes. 
Examining the dynamics of an internally coupled system that is driven by a constant thermodynamic force, we however find that the trade-off relation of a sub-system is not constrained by the minimal bound of conventional uncertainty relation. 
We made our point explicit by using an exactly solvable model of interacting oscillators.
As the number ($N$) of interacting oscillators increases, the uncertainty bound of individual oscillators is reduced to $2k_BT/N$ upon full synchronization under strong coupling. 
The cost-precision trade-off for the sub-system is particularly relevant for sub-cellular processes where collective dynamics emerges from multiple energy-expending components interacting with each other.
\end{abstract}

\maketitle
Orders that emerge in life are maintained via exchanges of energy, matters, and information with the environment~\cite{schrodinger1992, glansdorff1973}.
The recent advances in stochastic thermodynamics~\cite{qian2007, sekimoto2010, seifert2012} underscores the interplay between energy, information, and their trade-offs in small systems out of equilibrium, epitomized by diverse biological processes~\cite{perunov2016, england2013, cao2015, Barato2016, sartori2015, lan2012, sartori2015free, lang2014, goldt2017, ito2015}. 
To maintain a dissipative process at non-equilibrium steady states (NESS),
free energy consumption (or heat dissipation), $q(\tau)$, is bound to the precision of time-integrated output variable, $\theta(\tau)$, that characterizes the process ~\cite{cao2015, Barato2016}. 
The \emph{thermodynamic uncertainty relation} (TUR) concisely recapitulates this trade-off relation and specifies its minimal bound ~\cite{Barato2015}, 
\begin{equation}
\label{eq:tur}
{\cal Q} \equiv \langle q(\tau) \rangle \times \frac{\langle \delta \theta(\tau)^2 \rangle}{\langle \theta(\tau) \rangle^2} \ge \mathcal{Q}_{\text{min}}=2 k_B T,
\end{equation}
where $k_B$ is the Boltzmann constant, $T$ is temperature, and $\langle \cdots \rangle$ represents an average over an ensemble of many realizations.

Since the TUR being conjectured, general proofs and extension of the relation have been put forward from entirely different perspectives \cite{gingrich2016, shiraishi2016PRL,pietzonka2016PRE,pigolotti2017,pietzonka2017PRE}, which have greatly enriched our understanding of the dynamical processes in nonequilibrium.  
Further, analyses of experimental data exploiting the idea of TUR shine new light on problems associated with biological physics \cite{hwang2018JPCL}. 
It is remarkable that the thermodynamic consideration alone can provide a novel understanding into the design principle of molecular machines~\cite{bustamante2005,hwang2018JPCL}. 
Nonetheless, currently the focus of TUR is on the energetic cost for the operation of a single enzyme and molecular motor, or for a system  consisting of multiple energy-consuming modules as a whole.
In practice, measurement of energetic cost can be made on a sub-system that interacts with other modules of the whole system; that is, calculation of uncertainty measure $\mathcal{Q}$ can be made for the {\it sub-system}.  
For example, in biological cells, it is often recognized that the individual energy-consuming modules are not in strict isolation, but are coupled with others, displaying collective or cooperative dynamics \cite{julicher1995PRL,klumpp2005PNAS,Muller08PNAS,Um2012PRL}.

Studying the cost-precision trade-offs for internally coupled systems driven by a constant thermodynamic force, 
one could choose $\theta_i$, an output variable to monitor the time evolution of the $i$-th sub-system, and also measure $q_i$, the cost to operate the $i$-th sub-system. 
The uncertainty measure $\mathcal{Q}^{\text{sub}}={\cal Q}(q_i, \theta_i)$ which probes the trade-off between cost $q_i$ and squared relative error in the observable $\theta_i$, $\langle\delta\theta_i^2\rangle/\langle\theta_i\rangle^2$
could, in practice, be a more relevant measure to explore in studying sub-cellular processes than $\mathcal{Q}(q,\theta_i)$ which considers the cost of operating the whole system $q=\sum_iq_i$.

\begin{figure}
\centering
\includegraphics[width=7cm]{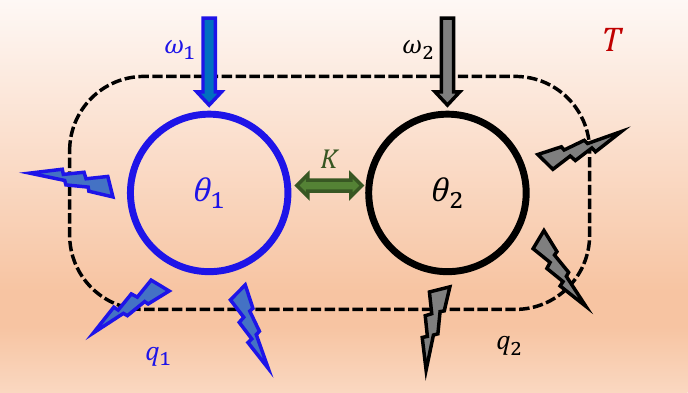}
\caption{(Color online) Two interacting oscillators embedded in a thermal reservoir with a temperature $T$. 
The motion of each oscillator, described by the phase variable $\theta_{1,2}$, is powered by the inherent frequency $\omega_{1,2}$, corresponding to the non-conservative driving force. 
The parameter $K$ characterizes the interaction strength between the oscillators, which elicits the synchronization of the phases. 
The heat dissipated from each oscillator is denoted by $q_{1,2}$. }
\label{fig1}
\end{figure}

In this Letter, we explore the cost-precision trade-off relation of a sub-system that is energetically coupled with the remaining part of the system as well as its thermal bath.
To this end, we investigate a concrete example, a system of interacting oscillators under thermal fluctuations (Fig.~\ref{fig1}).
When individual oscillators are non-interacting and independent from each other, the TUR still holds regardless of which output variable is probed and which part of cost is included for the calculation of uncertainty measure. 
We, however, show that when the dynamics of interacting oscillators are synchronized, 
the individual oscillator can achieve, with the same amount of energetic cost, a higher phase precision than the bound dictated by the conventional TUR.

\begin{figure}
\centering
\includegraphics[width=9cm]{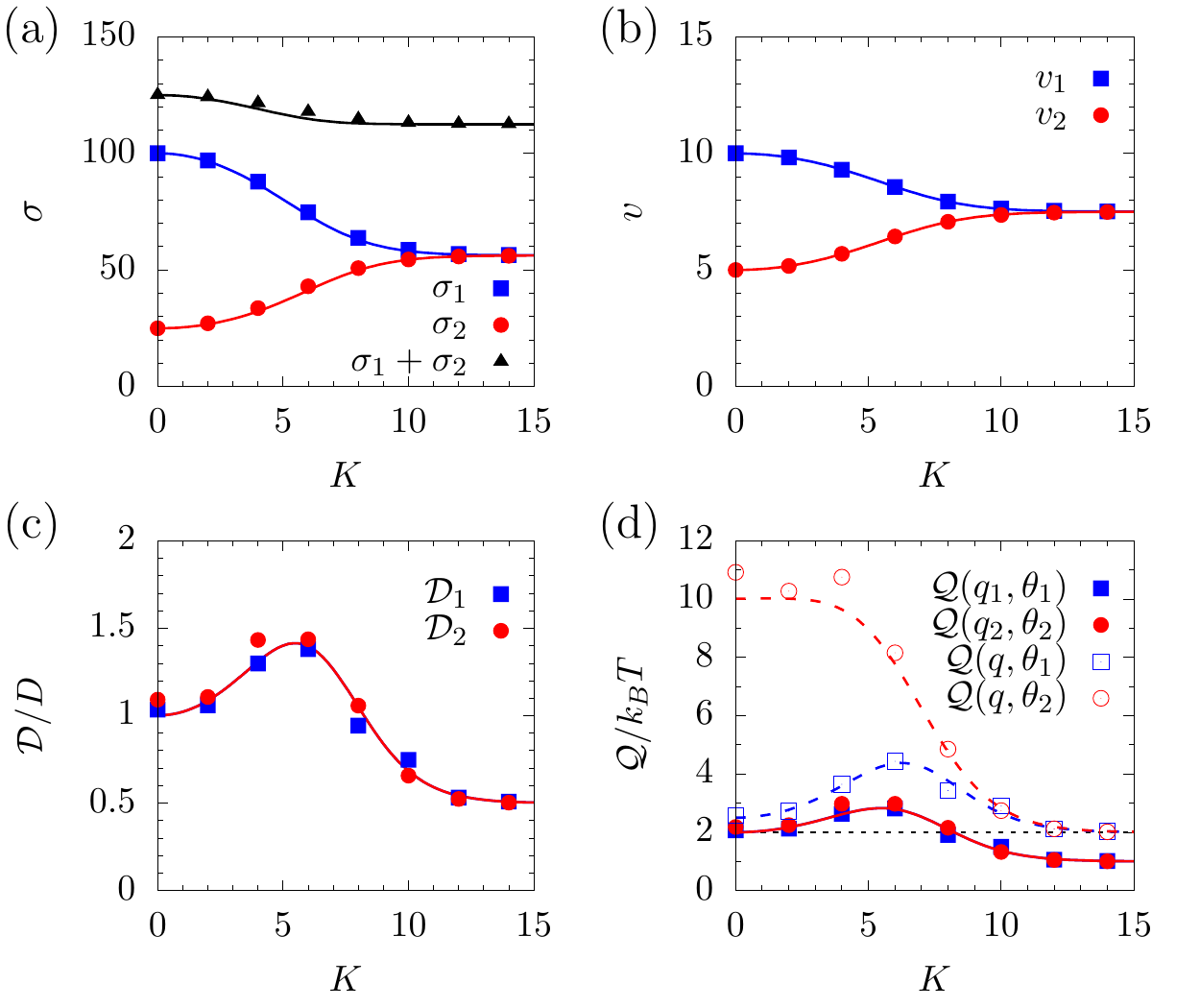}
\caption{(Color online) Thermodynamic uncertainty relations for two coupled oscillators.
(a) Heat dissipation rates of two oscillators (blue squares and red circles) and their total heat dissipation rate (black circles). 
The two oscillators have intrinsic frequencies of $\omega_1=10$ and $\omega_2=5$, respectively, and the noise strength is set to $D=1$.  
(b) Mean phase velocities of two oscillators.
(c) Relative diffusion constants that represent phase fluctuations.
(d) The uncertainty measures for the cost-precision trade-offs calculated for the total- ($\mathcal{Q}(q,\theta_i)$) and sub-systems ($\mathcal{Q}(q_i,\theta_i)$).
The dotted line represent the usual minimal bound of TUR, $2k_BT$.
To compute $\sigma$, $v$, and ${\cal D}$, an ensemble of $10^3$ realizations of stochastic process are used. 
The lines depict analytical expressions of $\sigma$, $v$, $\mathcal{D}$, and $\mathcal{Q}$~\cite{Note1}. }
\label{fig2}
\end{figure}

As our model system, we adopted the noisy Stuart-Landau oscillator that has recently been used to discuss the trade-off between the energetic cost and the precision of phase variable \cite{cao2015}.
First, the single noisy Stuart-Landau oscillator, in the absence of amplitude-phase coupling, indeed meets the minimal bound ${\cal Q}_{\text{min}}=2k_B T$ \footnote{See Supplemental Material at [URL will be inserted by publisher] for the derivation of the TUR for a Stuart-Landau oscillator; phase velocity and heat dissipation of coupled oscillators; and Figs. S1 and S2.}.
Next, to study the effect of coupling between two oscillators on the uncertainty relation, we consider the phase dynamics of noisy coupled oscillators~\cite{Acebron2005}:
\begin{eqnarray}
\frac{d\theta_1}{dt} &=& \omega_1 + \frac{K}{2} \sin(\theta_2 - \theta_1) + \eta_1, \nonumber\\
\frac{d\theta_2}{dt} &=& \omega_2 + \frac{K}{2} \sin(\theta_1 - \theta_2) + \eta_2,
\label{Kuramoto}
\end{eqnarray}
where $\omega_{1,2}$ are the intrinsic frequencies of the two oscillators, $K$ is the coupling strength, and $\eta_{1,2}$ are thermal fluctuations that satisfy $\langle \eta_i (t) \rangle =0$ and $\langle \eta_i(t) \eta_j(s) \rangle=2D \delta_{ij} \delta(t-s)$. 
Here, the noise represents the effect of the thermal bath, whose strength satisfies the Einstein relation, $D=\mu k_B T$. 
If $\omega_i \ll K$ and $\sqrt{D}$, a condition corresponding to the XY model, strong interaction between the oscillators or large noise suppress the regular oscillation~\cite{sasa2015}.
To impose the NESS condition without phase locking, we mainly consider the parameter range: $|\omega_i| > |K| \gg \sqrt{D}$.
Mapped on a Langevin equation, $\dot{\theta}_i = \mu F_i + \eta_i$, where $\mu$ is the motility coefficient, which we set to $\mu=1$ for convenience. 
The force term $F_i$ of the model is divided into the \emph{dissipative} ($\omega_i$) and \emph{conservative} forces ($f_{ji}=(K/2)\sin{(\theta_j-\theta_i)}$). 
Two points are of note: (i) the dissipative forces corresponding to $\omega_1$ and $\omega_2$ are the source of driving; (ii) $f_{12}+f_{21}=0$ so that the net contribution of the conservative force to the whole system is zero. 
When the steady state is reached, the heat generated from the oscillator is calculated using~\cite{sekimoto2010}
\begin{equation}
q_i(\tau) = \int_0^\tau dt F_i(t) \circ \frac{d\theta_i}{dt},
\end{equation}
where the notation $\circ$ indicates that the integral was taken in the Stratonovich sense. 
To examine the uncertainty relation of this system, we calculate the following three quantities:
\begin{align}
{\sigma_i} &= \lim_{\tau \to \infty} \frac{\langle q_i (\tau) \rangle}{\tau},\nonumber\\
v_i &= \lim_{\tau \to \infty} \frac{\langle \theta_i (\tau) \rangle}{\tau}, \nonumber\\
 {\cal D}_i &=  \lim_{\tau \to \infty} \frac{\langle \delta \theta_i^2 (\tau) \rangle}{2\tau},
\label{eq:D}
\end{align}
where $\sigma_i$, $v_i$, and ${\cal D}_i$ are the mean heat dissipation rate, mean phase velocity, and effective diffusion constant of the $i$-th oscillator, respectively. 
Integrating Eq.~(\ref{Kuramoto}) numerically, we obtain the three quantities as a function of coupling strength $K$.  
The two interacting oscillator model is, in fact, exactly solvable; thus, the results from the numerics (data points in Fig. \ref{fig2}) can also be compared with the analytical expressions~\cite{Note1} (lines in Fig.~\ref{fig2}) with no tuning parameter. A few points are worth noting: 

(i) In the absence of coupling, the energetic cost to operate each oscillator $\sigma_1$ and $\sigma_2$ depends on each of the driving frequency $\omega_1$ and $\omega_2$.
However, upon synchronization at large $K$, which slows down the oscillator 1 and speeds up the oscillator 2, converging the mean phase velocities of the two oscillators ($v_1$ and $v_2$) (Fig.~\ref{fig2}b), the amount of cost $\sigma_1$ and $\sigma_2$ becomes identical (Fig.~\ref{fig2}a).
Remarkably, the total energetic cost $\sigma_1+\sigma_2$ to operate the two oscillators under synchronization is smaller than that under small $K$~\cite{izumida2016}.


(ii) In the absence of coupling ($K=0$), the diffusivity of phase variable is determined by the noise strength of the thermal bath (${\cal D}_i = D$). Weak coupling (small $K$) elicits additional fluctuations in the phase variable (Eq.~(\ref{Kuramoto}), which gives rise to an increase in the diffusivity (${\cal D}_i > D$).
Stronger couplings (large $K$) engendering the phase synchronization reduce the phase fluctuation below the thermal noise strength (${\cal D}_i \to D/2$).
The effective diffusivity $\mathcal{D}_i$ is maximized at $K\approx|\omega_1-\omega_2|$ (Fig.~\ref{fig2}c).

(iii) Combining the three quantities, we evaluate the cost-precision trade-offs: 
${\cal Q}(q, \theta_i) = (\sum_j \sigma_j) (2 {\cal D}_i)/v_i^2$ for the whole system and 
${\cal Q}(q_i, \theta_i) = \sigma_i (2 {\cal D}_i)/v_i^2$ for the sub-systems.
In the absence of coupling ($K=0$), the sub-systems behave independently from each other. 
The uncertainty measure calculated for the heat dissipation and phase fluctuations of a sub-system attains the minimal bound ${\cal Q}(q_i, \theta_i) =2k_BT$ at $K=0$.
The uncertainty measure accounting for the whole heat dissipation is always greater than the minimal bound of the conventional TUR, i.e.,  ${\cal Q}(q, \theta_i) \ge 2k_BT$ (Fig.~\ref{fig2}d)  \cite{Barato2015}. 
In marked contrast, once the sub-systems are synchronized under strong coupling ($K\gg |\omega_1-\omega_2|$), 
the minimal bound of the uncertainty measure is reduced to ${\cal Q}(q_i, \theta_i) =k_BT$.
This implies that without demanding extra thermodynamic cost the interacting oscillators under constant thermodynamic driving can attain a higher precision in the phase variable via synchronization. 
In the presence of synergetic interactions between the components of the system, the bound of cost-precision trade-off relation germane to the sub-system can break the minimal bound of TUR.
This is the principal finding of this study.

In fact, the problem of two coupled oscillators is an exactly solvable problem. 
Analyses at limiting cases allow us to gain a better understanding of the trade-off relation of interacting sub-systems. 
A joint probability density $P(\theta_1, \theta_2, t)$ for the noisy coupled oscillators obeys the Fokker-Planck equation,
$\partial_t P=-\partial_{\theta_1} J_1 -\partial_{\theta_2} J_2$, 
where $J_i = F_i P - D\partial P/\partial{\theta_i}$ is a probability current.
The steady state condition $\partial P/\partial t = 0$ defines the steady state current $J_i^{ss}$.
For a given $J_i^{ss}$, which allows us to calculate $v_i = \int_0^{2\pi} d\theta_1 \int_0^{2\pi} d\theta_2 J_i^{ss}$
and $\sigma_i = \int_0^{2\pi} d\theta_1 \int_0^{2\pi} d\theta_2 F_i J_i^{ss}$.
The steady state current for each oscillator $J_i^{ss}$ is conveniently calculated by using an orthogonal coordinate, $\phi_1=\theta_1+\theta_2$ and $\phi_2=\theta_1-\theta_2$.
From Eq.~(\ref{Kuramoto}), the time-evolutions of the orthogonal coordinates are given as  
\begin{eqnarray}
\label{eq:phi1}
\frac{d \phi_1}{dt} &=& 2 \bar{\omega} + \xi_1 \\
\label{eq:phi2}
\frac{d \phi_2}{dt} &=& \Delta \omega - K \sin \phi_2 + \xi_2,
\end{eqnarray}
where $2 \bar{\omega}\equiv \omega_1+\omega_2$ and $\Delta \omega\equiv  \omega_1 - \omega_2$ and $\langle \xi_i(t) \rangle = 0$ and $\langle \xi_i(t) \xi_j(s) \rangle = 4D \delta_{ij} \delta(t-s)$. 
Note that Eqs.~(\ref{eq:phi1}) and (\ref{eq:phi2}) are isomorphic to the Brownian motion in tilted potentials, which have recently been used to study the TUR \cite{Hyeon2017PRE}, where $\mathcal{Q}$ was shown non-monotonic with the \emph{tilt} and its minimal bound was attained at both weak and strong tilt limits.
Note that the present result of non-monotonic dependence of $\mathcal{Q}$ with vayring $K$, demonstrated in Fig.~\ref{fig2}d, is realized in the large tilt limit where the force terms, $2\bar{\omega}$ and $\Delta \omega-K\sin{\phi_2}$, in Eqs.~(\ref{eq:phi1}) and (\ref{eq:phi2}) are greater than the thermal noise.  

Equation~(\ref{eq:phi1}) straightforwardly leads to $\langle \delta \phi_1^2(\tau) \rangle = 4D\tau$.
The phase fluctuation $\langle \delta \phi_2^2(\tau) \rangle = 4D_{\text{eff}} \tau$ under the tilted period potential $V(\phi_2) = -K \cos \phi_2 - \Delta \omega\times \phi_2$ is explicitly calculated using~\cite{reimann2002}
\begin{equation}
\frac{D_{\text{eff}}}{D}= \frac{\int_0^{2\pi} d\phi_2 I_{\mp}(\phi_2) I_+(\phi_2) I_-(\phi_2)}{\big[ \int_0^{2\pi} d\phi_2 I_{\mp}(\phi_2)\big]^3},
\end{equation}
where $I_+(\phi_2) =\exp[V(\phi_2)/2D] \int_{\phi_2 - 2\pi}^{\phi_2} d\psi \exp[-V(\psi)/2D]$ and
$I_-(\phi_2) =\exp[-V(\phi_2)/2D] \int^{\phi_2 + 2\pi}_{\phi_2} d\psi \exp[V(\psi)/2D]$.
The orthogonality condition $\langle\delta\phi_1\delta\phi_2\rangle=0$ leads to    
$\langle \delta \theta_1^2 \rangle = \langle \delta \theta_2^2 \rangle = ({\langle \delta \phi_1^2 \rangle + \langle \delta \phi_2^2 \rangle})/{4}$; thus  
the phase fluctuations of each oscillator are straightforwardly related to the diffusion constants obtained for the two orthogonal coordinates as
\begin{equation}
\langle \delta \theta_i^2(\tau) \rangle = 2 {\cal D}_i \tau= (D+D_{\text{eff}}) \tau.
\end{equation}
In the large $K$ limit, the oscillators are synchronized with a negligible phase difference ($\phi_2 = \theta_1 - \theta_2 \approx 0$), which linearizes Eq.~(\ref{eq:phi2}) to 
$d\phi_2/dt = \Delta \omega - K \phi_2 + \xi_2$.
The variance of $\phi_2$ from its formal solution 
is 
\begin{equation}
\langle \delta \phi_2^2 \rangle = \frac{2D}{|K|} \bigg[ 1 - \exp(-2|K|\tau) \bigg].  
\end{equation}
Thus, $\langle \delta \phi_2^2 \rangle = 0$ for $K\gg \Delta \omega$, and $\langle \delta \phi_2^2 \rangle = 4D \tau$ for  $K \to 0$.

Next, at the two limiting cases of $K\rightarrow 0$ and $K\gg \Delta \omega$, the rate of heat dissipation is the square of the mean phase velocity ($\sigma_i=v_i^2$):  
(i) For weak coupling limit ($K\rightarrow 0$), 
the two oscillators behave independently from each other and oscillate with their own phase velocities $v_i = \omega_i$, which leads to 
\begin{align}
\sigma_i = \omega_i^2;  
\end{align} 
(ii) For strong coupling limit ($K\gg \Delta \omega$), the motion of oscillators is synchronized with the phase velocity of $v_1 =v_2=(\omega_1+\omega_2)/2$, which gives rise to the heat dissipation of 
\begin{align}
\sigma_1=\sigma_2 = \frac{(\omega_1+\omega_2)^2}{4}. 
\end{align}

Finally, combining $v_i$, $\mathcal{D}_i$, and $\sigma_i$ to calculate ${\cal Q}(q_i, \theta_i) = \sigma_i(2{\cal D}_i)/v_i^2$, 
we confirm ${\cal Q}(q_i, \theta_i) = 2k_BT$ under weak coupling (small $K$),
whereas ${\cal Q}(q_i, \theta_i) = k_BT$ under strong coupling (large $K$).
Note that the reduced minimal bound of cost-precision trade-off uncertainty measure ${\cal Q}(q_i, \theta_i)$ under strong coupling 
is the outcome of reduction in phase fluctuations rather than the changes in heat dissipation or phase velocity.

\begin{figure}
\centering
\includegraphics[width=9cm]{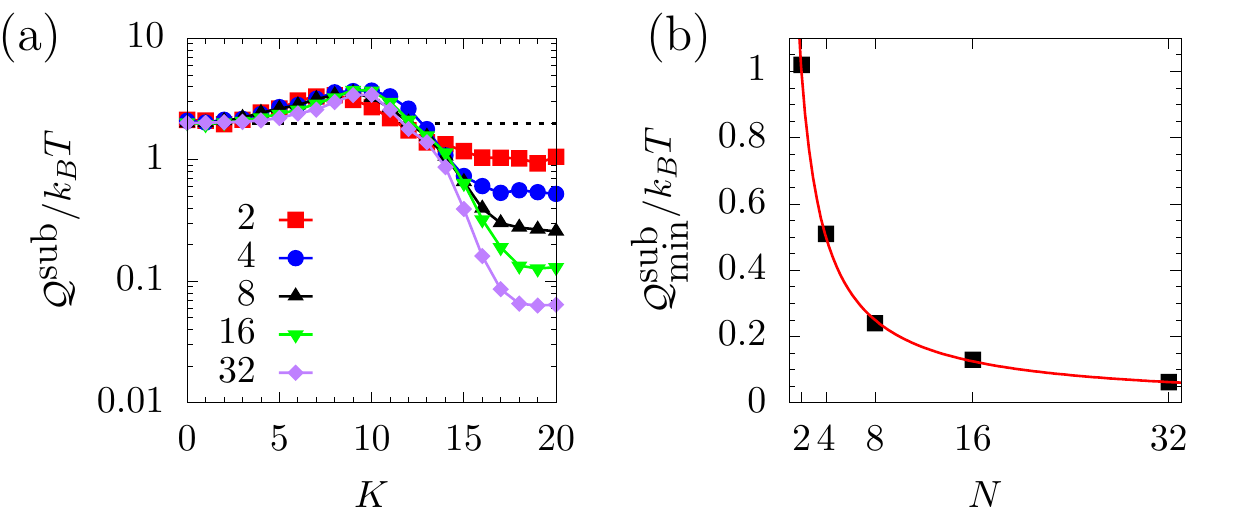}
\caption{(Color online) Thermodynamic uncertainty relations for sub-systems of interacting oscillators.
(a) $\mathcal{Q}^{\text{sub}}/k_BT$ for various numbers ($N=2, 4, 8, 16, 32$) of oscillators as a function of interaction strength $K$. 
Plotted are $\mathcal{Q}^{\text{sub}}/k_BT$ averaged over $N$ oscillators, 
Individual oscillators have intrinsic frequencies sampled from a normal distribution $\mathcal{N}(\bar{\omega},\Delta \omega)$ with a mean $\bar{\omega} = 10$ and a standard deviation $\Delta \omega=5$.
The noise strength is set to $D=1$. 
(b) The minimal bound of $\mathcal{Q}^{\text{sub}}$ for large $K \gg \Delta \omega$ ($\mathcal{Q}^{\text{sub}}_{\text{min}}$) for varying $N$ (data point) is obtained from the simulations of $N$-oscillator system. 
The line depicts $2/N$.}
\label{fig3}
\end{figure}

We generalize Eq.~(\ref{Kuramoto}) into $N$-coupled oscillators and explore how the bound of cost-precision relation for sub-systems changes with $N$. 
\begin{equation}
\label{eq:N}
\frac{d \theta_i}{dt} = \omega_i + \frac{K}{N} \sum_{j=1}^N \sin(\theta_j - \theta_i) + \eta_i,
\end{equation}
where we assume the driving frequency for the $i$-th oscillator $\omega_i$ sampled from a gaussian distribution with a mean $\bar{\omega}$ and a standard deviation $\Delta \omega$.
The numerically calculated ${\cal Q}(q_i, \theta_i)$ for the $N$-interacting oscillators confirms that the bound of TUR for sub-system is lowered by $1/N$ upon full synchronization (Fig.~\ref{fig3}). 
From the numerics, we find that the minimal bound of the uncertainty measure $\mathcal{Q}^{\text{sub}}$ for the individual oscillators for large $K$ in the $N$-interacting system inversely scales with $N$ as 
\begin{equation}
{\cal Q}^{\text{sub}}=\langle q_i(\tau)\rangle\times\frac{\langle\delta\theta(\tau)^2\rangle}{\langle \theta(\tau)\rangle^2}\geq  {\cal Q}^{\text{sub}}_{\text{min}} = \frac{2 k_BT}{N}.
\end{equation}
This finding can be rationalized with ease considering the following argument. 
In the limit of full synchronization, $\theta_i \approx \theta_j$; hence Eq.~(\ref{eq:N}) can be approximated as 
$d\theta_i/dt = \omega_i + \sum_j M_{ij} \theta_j + \eta_i$ with the interaction matrix
\begin{align}
{\bf M}=\begin{bmatrix}
    -K & \frac{K}{N} & \frac{K}{N} & \dots  & \frac{K}{N} \\
    \frac{K}{N} & -K & \frac{K}{N} & \dots  & \frac{K}{N} \\
    \vdots & \vdots & \vdots & \ddots & \vdots \\
    \frac{K}{N} & \frac{K}{N} & \frac{K}{N} & \dots  & -K
\end{bmatrix}.
\end{align}
{\bf M} has one zero eigenvalue ($\lambda_1=0$) with corresponding eigenvector $\phi_1 = \theta_1+\theta_2 + \cdots + \theta_N$.
The other $(N-1)$ eigenvalues are all negative with corresponding eigenvectors, $\phi_i = \theta_1-\theta_i$ for $i \in \{2, 3, \cdots, N\}$.
Then, using $\theta_1= (\phi_1 + \phi_2 + \cdots + \phi_N)/N$, one can obtain
\begin{equation}
\label{eq:phasefluctuation}
\langle \delta \theta_1^2 \rangle = \frac{\sum_{i=1}^{N} \langle \delta \phi_i^2 \rangle}{N^2} \approx \frac{\langle \delta \phi_1^2 \rangle}{N^2}= \frac{2D\tau}{N},
\end{equation}
where we have used $\langle \delta \phi_i^2 \rangle \to 0$ for $i \neq 1$ at steady states ($\tau\rightarrow \infty$) because their eigenvalues are negative.
Furthermore, in the limit of phase synchrony ($\phi_i = \theta_1 - \theta_i \approx 0$), the phase fluctuations are equivalent $\langle \delta \theta_i^2 \rangle = \langle \delta \theta_1^2 \rangle$ for all $i$.
Since the phase synchrony leads to $\langle q_i \rangle /\tau = \langle \theta_i \rangle^2$, every sub-system achieves the reduced lower bound ${\cal Q}^{\text{sub}}_{\text{min}} = {2 k_BT}/{N}$ with the reduced phase fluctuation in Eq.~(\ref{eq:phasefluctuation}).

Whereas the TUR sets a minimal bound ($\mathcal{Q}_{\text{min}}=2k_BT$) to the energy cost to operate an entire system for a given precision in the output observable,
many experimental measurements are often made by focusing on the dynamic process of sub-systems.  
For a system of interacting oscillators, we demonstrated that when the coupling between the oscillators is strong enough to elicit a full synchronization, individual oscillators could achieve a higher phase precision with a reduced net energetic cost.
From the perspective of TUR it is remarkable that the phase precision of the individual oscillator is benefitted by the interaction with other oscillators without incurring extra energetic cost. 
It has recently been pointed out that whereas a system represented by biochemical clock driven by a fixed thermodynamic force satisfies the TUR, which corresponds to verifying $\mathcal{Q}(q,\theta_i)\ge 2k_BT$ in this study, stochastic clocks driven by a periodic external protocol can achieve arbitrary precision at arbitrarily low cost~\cite{Barato2016}. 
The scenario that the precision increased by the mutual interaction between sub-systems under constant thermodynamic force has not been considered before; yet is deemed quite relevant to the experimental situation encountered in sub-cellular systems \cite{gross2007CurrBiol}.   
Here, the mutual interaction, which leads to the synchronization, is conservative in nature that the net force exchanged between sub-systems is compensated to be zero.
In general, any periodic coupling with odd parity contributes to a higher phase precision as long as the coupling is sufficiently strong~\cite{Note1}.

We note that the in-phase synchrony of interacting oscillators is not a prerequisite for $\mathcal{Q}_{\text{min}}^{\text{sub}}\leq \mathcal{Q}_{\text{min}}(=2k_BT)$. 
Other states that result from collective dynamics, such as states generated by out-of-phase synchrony, splay states of repulsive oscillators, and cluster states of the mixture of attractive and repulsive oscillators, can also reduce the energetic cost below $\mathcal{Q}_{\text{min}}$~\cite{Note1}.

The synchronization between interacting oscillators not only reduces total energetic cost to operate the whole system~\cite{izumida2016}, but also enhances the efficiency of power transfer between oscillators~\cite{imparato2015}.
The key finding underscored in this study, the reduction of the bound of cost-precision trade-off for sub-systems through synchronization, sheds light on the design principle underlying biological systems. 
Cooperative cargo transport by multiple motors~\cite{klumpp2005PNAS}, force generation by the coordination among muscle proteins~\cite{julicher1995PRL}, and synchronization of bacterial flagellar motors via hydrodynamic coupling~\cite{reichert2005EPJE,golestanian2011SoftMatter,marchetti2013RMP} are the seminal examples that the whole system divided into multiple energy-expending modules improves the precision of biological function by means of synchronization. 

To conclude, our generalized uncertainty relation offers a valuable insight into the trade-off principle underlying biological processes that have many interacting components. It would also of great interest to survey the diverse forms of collective dynamics \cite{Strogatz2017NatCommun} from the perspective of TUR. 


{\it Acknowledgement.}
This research was supported by the Basic Science Research Program through the National Research Foundation of Korea (NRF), funded by the Ministry of Education (2016R1D1A1B03932264) (J.J.). 
We acknowledge the Center for Advanced Computation in KIAS for providing computing resources.

\bibliography{uncertainty,mybib1} 

\begin{thebibliography}{39}%
\makeatletter
\providecommand \@ifxundefined [1]{%
 \@ifx{#1\undefined}
}%
\providecommand \@ifnum [1]{%
 \ifnum #1\expandafter \@firstoftwo
 \else \expandafter \@secondoftwo
 \fi
}%
\providecommand \@ifx [1]{%
 \ifx #1\expandafter \@firstoftwo
 \else \expandafter \@secondoftwo
 \fi
}%
\providecommand \natexlab [1]{#1}%
\providecommand \enquote  [1]{``#1''}%
\providecommand \bibnamefont  [1]{#1}%
\providecommand \bibfnamefont [1]{#1}%
\providecommand \citenamefont [1]{#1}%
\providecommand \href@noop [0]{\@secondoftwo}%
\providecommand \href [0]{\begingroup \@sanitize@url \@href}%
\providecommand \@href[1]{\@@startlink{#1}\@@href}%
\providecommand \@@href[1]{\endgroup#1\@@endlink}%
\providecommand \@sanitize@url [0]{\catcode `\\12\catcode `\$12\catcode
  `\&12\catcode `\#12\catcode `\^12\catcode `\_12\catcode `\%12\relax}%
\providecommand \@@startlink[1]{}%
\providecommand \@@endlink[0]{}%
\providecommand \url  [0]{\begingroup\@sanitize@url \@url }%
\providecommand \@url [1]{\endgroup\@href {#1}{\urlprefix }}%
\providecommand \urlprefix  [0]{URL }%
\providecommand \Eprint [0]{\href }%
\providecommand \doibase [0]{http://dx.doi.org/}%
\providecommand \selectlanguage [0]{\@gobble}%
\providecommand \bibinfo  [0]{\@secondoftwo}%
\providecommand \bibfield  [0]{\@secondoftwo}%
\providecommand \translation [1]{[#1]}%
\providecommand \BibitemOpen [0]{}%
\providecommand \bibitemStop [0]{}%
\providecommand \bibitemNoStop [0]{.\EOS\space}%
\providecommand \EOS [0]{\spacefactor3000\relax}%
\providecommand \BibitemShut  [1]{\csname bibitem#1\endcsname}%
\let\auto@bib@innerbib\@empty
\bibitem [{\citenamefont {Schr{\"o}dinger}(1992)}]{schrodinger1992}%
  \BibitemOpen
  \bibfield  {author} {\bibinfo {author} {\bibfnamefont {E.}~\bibnamefont
  {Schr{\"o}dinger}},\ }\href@noop {} {\emph {\bibinfo {title} {What is life?:
  With mind and matter and autobiographical sketches}}}\ (\bibinfo  {publisher}
  {Cambridge University Press},\ \bibinfo {year} {1992})\BibitemShut {NoStop}%
\bibitem [{\citenamefont {Glansdorff}\ and\ \citenamefont
  {Prigogine}(1973)}]{glansdorff1973}%
  \BibitemOpen
  \bibfield  {author} {\bibinfo {author} {\bibfnamefont {P.}~\bibnamefont
  {Glansdorff}}\ and\ \bibinfo {author} {\bibfnamefont {I.}~\bibnamefont
  {Prigogine}},\ }\href@noop {} {\bibfield  {journal} {\bibinfo  {journal}
  {Amer. J. Phys.}\ }\textbf {\bibinfo {volume} {41}},\ \bibinfo {pages} {147}
  (\bibinfo {year} {1973})}\BibitemShut {NoStop}%
\bibitem [{\citenamefont {Qian}(2007)}]{qian2007}%
  \BibitemOpen
  \bibfield  {author} {\bibinfo {author} {\bibfnamefont {H.}~\bibnamefont
  {Qian}},\ }\href@noop {} {\bibfield  {journal} {\bibinfo  {journal} {Annu.
  Rev. Phys. Chem.}\ }\textbf {\bibinfo {volume} {58}},\ \bibinfo {pages} {113}
  (\bibinfo {year} {2007})}\BibitemShut {NoStop}%
\bibitem [{\citenamefont {Sekimoto}(2010)}]{sekimoto2010}%
  \BibitemOpen
  \bibfield  {author} {\bibinfo {author} {\bibfnamefont {K.}~\bibnamefont
  {Sekimoto}},\ }\href@noop {} {\emph {\bibinfo {title} {Stochastic
  energetics}}},\ Vol.\ \bibinfo {volume} {799}\ (\bibinfo  {publisher}
  {Springer},\ \bibinfo {year} {2010})\BibitemShut {NoStop}%
\bibitem [{\citenamefont {Seifert}(2012)}]{seifert2012}%
  \BibitemOpen
  \bibfield  {author} {\bibinfo {author} {\bibfnamefont {U.}~\bibnamefont
  {Seifert}},\ }\href@noop {} {\bibfield  {journal} {\bibinfo  {journal}
  {Reports on Progress in Physics}\ }\textbf {\bibinfo {volume} {75}},\
  \bibinfo {pages} {126001} (\bibinfo {year} {2012})}\BibitemShut {NoStop}%
\bibitem [{\citenamefont {Perunov}\ \emph {et~al.}(2016)\citenamefont
  {Perunov}, \citenamefont {Marsland},\ and\ \citenamefont
  {England}}]{perunov2016}%
  \BibitemOpen
  \bibfield  {author} {\bibinfo {author} {\bibfnamefont {N.}~\bibnamefont
  {Perunov}}, \bibinfo {author} {\bibfnamefont {R.~A.}\ \bibnamefont
  {Marsland}}, \ and\ \bibinfo {author} {\bibfnamefont {J.~L.}\ \bibnamefont
  {England}},\ }\href@noop {} {\bibfield  {journal} {\bibinfo  {journal} {Phys.
  Rev. X}\ }\textbf {\bibinfo {volume} {6}},\ \bibinfo {pages} {021036}
  (\bibinfo {year} {2016})}\BibitemShut {NoStop}%
\bibitem [{\citenamefont {England}(2013)}]{england2013}%
  \BibitemOpen
  \bibfield  {author} {\bibinfo {author} {\bibfnamefont {J.~L.}\ \bibnamefont
  {England}},\ }\href@noop {} {\bibfield  {journal} {\bibinfo  {journal} {J.
  Chem. Phys.}\ }\textbf {\bibinfo {volume} {139}},\ \bibinfo {pages}
  {09B623\_1} (\bibinfo {year} {2013})}\BibitemShut {NoStop}%
\bibitem [{\citenamefont {Cao}\ \emph {et~al.}(2015)\citenamefont {Cao},
  \citenamefont {Wang}, \citenamefont {Ouyang},\ and\ \citenamefont
  {Tu}}]{cao2015}%
  \BibitemOpen
  \bibfield  {author} {\bibinfo {author} {\bibfnamefont {Y.}~\bibnamefont
  {Cao}}, \bibinfo {author} {\bibfnamefont {H.}~\bibnamefont {Wang}}, \bibinfo
  {author} {\bibfnamefont {Q.}~\bibnamefont {Ouyang}}, \ and\ \bibinfo {author}
  {\bibfnamefont {Y.}~\bibnamefont {Tu}},\ }\href@noop {} {\bibfield  {journal}
  {\bibinfo  {journal} {Nature Physics}\ }\textbf {\bibinfo {volume} {11}},\
  \bibinfo {pages} {772} (\bibinfo {year} {2015})}\BibitemShut {NoStop}%
\bibitem [{\citenamefont {Barato}\ and\ \citenamefont
  {Seifert}(2016)}]{Barato2016}%
  \BibitemOpen
  \bibfield  {author} {\bibinfo {author} {\bibfnamefont {A.~C.}\ \bibnamefont
  {Barato}}\ and\ \bibinfo {author} {\bibfnamefont {U.}~\bibnamefont
  {Seifert}},\ }\href@noop {} {\bibfield  {journal} {\bibinfo  {journal} {Phys.
  Rev. X}\ }\textbf {\bibinfo {volume} {6}},\ \bibinfo {pages} {041053}
  (\bibinfo {year} {2016})}\BibitemShut {NoStop}%
\bibitem [{\citenamefont {Sartori}\ and\ \citenamefont
  {Pigolotti}(2015)}]{sartori2015}%
  \BibitemOpen
  \bibfield  {author} {\bibinfo {author} {\bibfnamefont {P.}~\bibnamefont
  {Sartori}}\ and\ \bibinfo {author} {\bibfnamefont {S.}~\bibnamefont
  {Pigolotti}},\ }\href@noop {} {\bibfield  {journal} {\bibinfo  {journal}
  {Phys. Rev. X}\ }\textbf {\bibinfo {volume} {5}},\ \bibinfo {pages} {041039}
  (\bibinfo {year} {2015})}\BibitemShut {NoStop}%
\bibitem [{\citenamefont {Lan}\ \emph {et~al.}(2012)\citenamefont {Lan},
  \citenamefont {Sartori}, \citenamefont {Neumann}, \citenamefont {Sourjik},\
  and\ \citenamefont {Tu}}]{lan2012}%
  \BibitemOpen
  \bibfield  {author} {\bibinfo {author} {\bibfnamefont {G.}~\bibnamefont
  {Lan}}, \bibinfo {author} {\bibfnamefont {P.}~\bibnamefont {Sartori}},
  \bibinfo {author} {\bibfnamefont {S.}~\bibnamefont {Neumann}}, \bibinfo
  {author} {\bibfnamefont {V.}~\bibnamefont {Sourjik}}, \ and\ \bibinfo
  {author} {\bibfnamefont {Y.}~\bibnamefont {Tu}},\ }\href@noop {} {\bibfield
  {journal} {\bibinfo  {journal} {Nature Physics}\ }\textbf {\bibinfo {volume}
  {8}},\ \bibinfo {pages} {422} (\bibinfo {year} {2012})}\BibitemShut {NoStop}%
\bibitem [{\citenamefont {Sartori}\ and\ \citenamefont
  {Tu}(2015)}]{sartori2015free}%
  \BibitemOpen
  \bibfield  {author} {\bibinfo {author} {\bibfnamefont {P.}~\bibnamefont
  {Sartori}}\ and\ \bibinfo {author} {\bibfnamefont {Y.}~\bibnamefont {Tu}},\
  }\href@noop {} {\bibfield  {journal} {\bibinfo  {journal} {Phys. Rev. Lett.}\
  }\textbf {\bibinfo {volume} {115}},\ \bibinfo {pages} {118102} (\bibinfo
  {year} {2015})}\BibitemShut {NoStop}%
\bibitem [{\citenamefont {Lang}\ \emph {et~al.}(2014)\citenamefont {Lang},
  \citenamefont {Fisher}, \citenamefont {Mora},\ and\ \citenamefont
  {Mehta}}]{lang2014}%
  \BibitemOpen
  \bibfield  {author} {\bibinfo {author} {\bibfnamefont {A.~H.}\ \bibnamefont
  {Lang}}, \bibinfo {author} {\bibfnamefont {C.~K.}\ \bibnamefont {Fisher}},
  \bibinfo {author} {\bibfnamefont {T.}~\bibnamefont {Mora}}, \ and\ \bibinfo
  {author} {\bibfnamefont {P.}~\bibnamefont {Mehta}},\ }\href@noop {}
  {\bibfield  {journal} {\bibinfo  {journal} {Phys. Rev. Lett.}\ }\textbf
  {\bibinfo {volume} {113}},\ \bibinfo {pages} {148103} (\bibinfo {year}
  {2014})}\BibitemShut {NoStop}%
\bibitem [{\citenamefont {Goldt}\ and\ \citenamefont
  {Seifert}(2017)}]{goldt2017}%
  \BibitemOpen
  \bibfield  {author} {\bibinfo {author} {\bibfnamefont {S.}~\bibnamefont
  {Goldt}}\ and\ \bibinfo {author} {\bibfnamefont {U.}~\bibnamefont
  {Seifert}},\ }\href@noop {} {\bibfield  {journal} {\bibinfo  {journal} {Phys.
  Rev. Lett.}\ }\textbf {\bibinfo {volume} {118}},\ \bibinfo {pages} {010601}
  (\bibinfo {year} {2017})}\BibitemShut {NoStop}%
\bibitem [{\citenamefont {Ito}\ and\ \citenamefont {Sagawa}(2015)}]{ito2015}%
  \BibitemOpen
  \bibfield  {author} {\bibinfo {author} {\bibfnamefont {S.}~\bibnamefont
  {Ito}}\ and\ \bibinfo {author} {\bibfnamefont {T.}~\bibnamefont {Sagawa}},\
  }\href@noop {} {\bibfield  {journal} {\bibinfo  {journal} {Nat. Commun.}\
  }\textbf {\bibinfo {volume} {6}},\ \bibinfo {pages} {7498} (\bibinfo {year}
  {2015})}\BibitemShut {NoStop}%
\bibitem [{\citenamefont {Barato}\ and\ \citenamefont
  {Seifert}(2015)}]{Barato2015}%
  \BibitemOpen
  \bibfield  {author} {\bibinfo {author} {\bibfnamefont {A.~C.}\ \bibnamefont
  {Barato}}\ and\ \bibinfo {author} {\bibfnamefont {U.}~\bibnamefont
  {Seifert}},\ }\href@noop {} {\bibfield  {journal} {\bibinfo  {journal} {Phys.
  Rev. Lett.}\ }\textbf {\bibinfo {volume} {114}},\ \bibinfo {pages} {158101}
  (\bibinfo {year} {2015})}\BibitemShut {NoStop}%
\bibitem [{\citenamefont {Gingrich}\ \emph {et~al.}(2016)\citenamefont
  {Gingrich}, \citenamefont {Horowitz}, \citenamefont {Perunov},\ and\
  \citenamefont {England}}]{gingrich2016}%
  \BibitemOpen
  \bibfield  {author} {\bibinfo {author} {\bibfnamefont {T.~R.}\ \bibnamefont
  {Gingrich}}, \bibinfo {author} {\bibfnamefont {J.~M.}\ \bibnamefont
  {Horowitz}}, \bibinfo {author} {\bibfnamefont {N.}~\bibnamefont {Perunov}}, \
  and\ \bibinfo {author} {\bibfnamefont {J.~L.}\ \bibnamefont {England}},\
  }\href@noop {} {\bibfield  {journal} {\bibinfo  {journal} {Phys. Rev. Lett.}\
  }\textbf {\bibinfo {volume} {116}},\ \bibinfo {pages} {120601} (\bibinfo
  {year} {2016})}\BibitemShut {NoStop}%
\bibitem [{\citenamefont {Shiraishi}\ \emph {et~al.}(2016)\citenamefont
  {Shiraishi}, \citenamefont {Saito},\ and\ \citenamefont
  {Tasaki}}]{shiraishi2016PRL}%
  \BibitemOpen
  \bibfield  {author} {\bibinfo {author} {\bibfnamefont {N.}~\bibnamefont
  {Shiraishi}}, \bibinfo {author} {\bibfnamefont {K.}~\bibnamefont {Saito}}, \
  and\ \bibinfo {author} {\bibfnamefont {H.}~\bibnamefont {Tasaki}},\
  }\href@noop {} {\bibfield  {journal} {\bibinfo  {journal} {Phys. Rev. Lett.}\
  }\textbf {\bibinfo {volume} {117}},\ \bibinfo {pages} {190601} (\bibinfo
  {year} {2016})}\BibitemShut {NoStop}%
\bibitem [{\citenamefont {Pietzonka}\ \emph {et~al.}(2016)\citenamefont
  {Pietzonka}, \citenamefont {Barato},\ and\ \citenamefont
  {Seifert}}]{pietzonka2016PRE}%
  \BibitemOpen
  \bibfield  {author} {\bibinfo {author} {\bibfnamefont {P.}~\bibnamefont
  {Pietzonka}}, \bibinfo {author} {\bibfnamefont {A.~C.}\ \bibnamefont
  {Barato}}, \ and\ \bibinfo {author} {\bibfnamefont {U.}~\bibnamefont
  {Seifert}},\ }\href@noop {} {\bibfield  {journal} {\bibinfo  {journal} {Phys.
  Rev. E.}\ }\textbf {\bibinfo {volume} {93}},\ \bibinfo {pages} {052145}
  (\bibinfo {year} {2016})}\BibitemShut {NoStop}%
\bibitem [{\citenamefont {Pigolotti}\ \emph {et~al.}(2017)\citenamefont
  {Pigolotti}, \citenamefont {Neri}, \citenamefont {Rold{\'a}n},\ and\
  \citenamefont {J{\"u}licher}}]{pigolotti2017}%
  \BibitemOpen
  \bibfield  {author} {\bibinfo {author} {\bibfnamefont {S.}~\bibnamefont
  {Pigolotti}}, \bibinfo {author} {\bibfnamefont {I.}~\bibnamefont {Neri}},
  \bibinfo {author} {\bibfnamefont {{\'E}.}~\bibnamefont {Rold{\'a}n}}, \ and\
  \bibinfo {author} {\bibfnamefont {F.}~\bibnamefont {J{\"u}licher}},\
  }\href@noop {} {\bibfield  {journal} {\bibinfo  {journal} {Phys. Rev. Lett.}\
  }\textbf {\bibinfo {volume} {119}},\ \bibinfo {pages} {140604} (\bibinfo
  {year} {2017})}\BibitemShut {NoStop}%
\bibitem [{\citenamefont {Pietzonka}\ \emph {et~al.}(2017)\citenamefont
  {Pietzonka}, \citenamefont {Ritort},\ and\ \citenamefont
  {Seifert}}]{pietzonka2017PRE}%
  \BibitemOpen
  \bibfield  {author} {\bibinfo {author} {\bibfnamefont {P.}~\bibnamefont
  {Pietzonka}}, \bibinfo {author} {\bibfnamefont {F.}~\bibnamefont {Ritort}}, \
  and\ \bibinfo {author} {\bibfnamefont {U.}~\bibnamefont {Seifert}},\
  }\href@noop {} {\bibfield  {journal} {\bibinfo  {journal} {Phys. Rev. E.}\
  }\textbf {\bibinfo {volume} {96}},\ \bibinfo {pages} {012101} (\bibinfo
  {year} {2017})}\BibitemShut {NoStop}%
\bibitem [{\citenamefont {Hwang}\ and\ \citenamefont
  {Hyeon}(2018)}]{hwang2018JPCL}%
  \BibitemOpen
  \bibfield  {author} {\bibinfo {author} {\bibfnamefont {W.}~\bibnamefont
  {Hwang}}\ and\ \bibinfo {author} {\bibfnamefont {C.}~\bibnamefont {Hyeon}},\
  }\href@noop {} {\bibfield  {journal} {\bibinfo  {journal} {J. Phys. Chem.
  Lett.}\ }\textbf {\bibinfo {volume} {9}},\ \bibinfo {pages} {513} (\bibinfo
  {year} {2018})}\BibitemShut {NoStop}%
\bibitem [{\citenamefont {Bustamante}\ \emph {et~al.}(2005)\citenamefont
  {Bustamante}, \citenamefont {Liphardt},\ and\ \citenamefont
  {Ritort}}]{bustamante2005}%
  \BibitemOpen
  \bibfield  {author} {\bibinfo {author} {\bibfnamefont {C.}~\bibnamefont
  {Bustamante}}, \bibinfo {author} {\bibfnamefont {J.}~\bibnamefont
  {Liphardt}}, \ and\ \bibinfo {author} {\bibfnamefont {F.}~\bibnamefont
  {Ritort}},\ }\href@noop {} {\bibfield  {journal} {\bibinfo  {journal}
  {Physics Today}\ }\textbf {\bibinfo {volume} {58}},\ \bibinfo {pages} {43}
  (\bibinfo {year} {2005})}\BibitemShut {NoStop}%
\bibitem [{\citenamefont {J{\"u}licher}\ and\ \citenamefont
  {Prost}(1995)}]{julicher1995PRL}%
  \BibitemOpen
  \bibfield  {author} {\bibinfo {author} {\bibfnamefont {F.}~\bibnamefont
  {J{\"u}licher}}\ and\ \bibinfo {author} {\bibfnamefont {J.}~\bibnamefont
  {Prost}},\ }\href@noop {} {\bibfield  {journal} {\bibinfo  {journal} {Phys.
  Rev. Lett.}\ }\textbf {\bibinfo {volume} {75}},\ \bibinfo {pages} {2618}
  (\bibinfo {year} {1995})}\BibitemShut {NoStop}%
\bibitem [{\citenamefont {Klumpp}\ and\ \citenamefont
  {Lipowsky}(2005)}]{klumpp2005PNAS}%
  \BibitemOpen
  \bibfield  {author} {\bibinfo {author} {\bibfnamefont {S.}~\bibnamefont
  {Klumpp}}\ and\ \bibinfo {author} {\bibfnamefont {R.}~\bibnamefont
  {Lipowsky}},\ }\href@noop {} {\bibfield  {journal} {\bibinfo  {journal}
  {Proc. Natl. Acad. Sci. U. S. A.}\ }\textbf {\bibinfo {volume} {102}},\
  \bibinfo {pages} {17284} (\bibinfo {year} {2005})}\BibitemShut {NoStop}%
\bibitem [{\citenamefont {Muller}\ \emph {et~al.}(2008)\citenamefont {Muller},
  \citenamefont {Klumpp},\ and\ \citenamefont {Lipowsky}}]{Muller08PNAS}%
  \BibitemOpen
  \bibfield  {author} {\bibinfo {author} {\bibfnamefont {M.~J.~I.}\
  \bibnamefont {Muller}}, \bibinfo {author} {\bibfnamefont {S.}~\bibnamefont
  {Klumpp}}, \ and\ \bibinfo {author} {\bibfnamefont {R.}~\bibnamefont
  {Lipowsky}},\ }\href@noop {} {\bibfield  {journal} {\bibinfo  {journal}
  {Proc. Natl. Acad. Sci. U. S. A.}\ }\textbf {\bibinfo {volume} {105}},\
  \bibinfo {pages} {4609} (\bibinfo {year} {2008})}\BibitemShut {NoStop}%
\bibitem [{\citenamefont {Um}\ \emph {et~al.}(2012)\citenamefont {Um},
  \citenamefont {Hong}, \citenamefont {Marchesoni},\ and\ \citenamefont
  {Park}}]{Um2012PRL}%
  \BibitemOpen
  \bibfield  {author} {\bibinfo {author} {\bibfnamefont {J.}~\bibnamefont
  {Um}}, \bibinfo {author} {\bibfnamefont {H.}~\bibnamefont {Hong}}, \bibinfo
  {author} {\bibfnamefont {F.}~\bibnamefont {Marchesoni}}, \ and\ \bibinfo
  {author} {\bibfnamefont {H.}~\bibnamefont {Park}},\ }\href@noop {} {\bibfield
   {journal} {\bibinfo  {journal} {Phys. Rev. Lett.}\ }\textbf {\bibinfo
  {volume} {108}},\ \bibinfo {pages} {060601} (\bibinfo {year}
  {2012})}\BibitemShut {NoStop}%
\bibitem [{Note1()}]{Note1}%
  \BibitemOpen
  \bibinfo {note} {See Supplemental Material at [URL will be inserted by
  publisher] for the derivation of the TUR for a Stuart-Landau oscillator;
  phase velocity and heat dissipation of coupled oscillators; and Figs. S1 and
  S2.}\BibitemShut {Stop}%
\bibitem [{\citenamefont {Acebr{\'o}n}\ \emph {et~al.}(2005)\citenamefont
  {Acebr{\'o}n}, \citenamefont {Bonilla}, \citenamefont {Vicente},
  \citenamefont {Ritort},\ and\ \citenamefont {Spigler}}]{Acebron2005}%
  \BibitemOpen
  \bibfield  {author} {\bibinfo {author} {\bibfnamefont {J.~A.}\ \bibnamefont
  {Acebr{\'o}n}}, \bibinfo {author} {\bibfnamefont {L.~L.}\ \bibnamefont
  {Bonilla}}, \bibinfo {author} {\bibfnamefont {C.~J.~P.}\ \bibnamefont
  {Vicente}}, \bibinfo {author} {\bibfnamefont {F.}~\bibnamefont {Ritort}}, \
  and\ \bibinfo {author} {\bibfnamefont {R.}~\bibnamefont {Spigler}},\
  }\href@noop {} {\bibfield  {journal} {\bibinfo  {journal} {Rev. Mod. Phys.}\
  }\textbf {\bibinfo {volume} {77}},\ \bibinfo {pages} {137} (\bibinfo {year}
  {2005})}\BibitemShut {NoStop}%
\bibitem [{\citenamefont {Sasa}(2015)}]{sasa2015}%
  \BibitemOpen
  \bibfield  {author} {\bibinfo {author} {\bibfnamefont {S.-I.}\ \bibnamefont
  {Sasa}},\ }\href@noop {} {\bibfield  {journal} {\bibinfo  {journal} {New J.
  Phys.}\ }\textbf {\bibinfo {volume} {17}},\ \bibinfo {pages} {045024}
  (\bibinfo {year} {2015})}\BibitemShut {NoStop}%
\bibitem [{\citenamefont {Izumida}\ \emph {et~al.}(2016)\citenamefont
  {Izumida}, \citenamefont {Kori},\ and\ \citenamefont
  {Seifert}}]{izumida2016}%
  \BibitemOpen
  \bibfield  {author} {\bibinfo {author} {\bibfnamefont {Y.}~\bibnamefont
  {Izumida}}, \bibinfo {author} {\bibfnamefont {H.}~\bibnamefont {Kori}}, \
  and\ \bibinfo {author} {\bibfnamefont {U.}~\bibnamefont {Seifert}},\
  }\href@noop {} {\bibfield  {journal} {\bibinfo  {journal} {Phys. Rev. E}\
  }\textbf {\bibinfo {volume} {94}},\ \bibinfo {pages} {052221} (\bibinfo
  {year} {2016})}\BibitemShut {NoStop}%
\bibitem [{\citenamefont {Hyeon}\ and\ \citenamefont
  {Hwang}(2017)}]{Hyeon2017PRE}%
  \BibitemOpen
  \bibfield  {author} {\bibinfo {author} {\bibfnamefont {C.}~\bibnamefont
  {Hyeon}}\ and\ \bibinfo {author} {\bibfnamefont {W.}~\bibnamefont {Hwang}},\
  }\href@noop {} {\bibfield  {journal} {\bibinfo  {journal} {Phys. Rev. E.}\
  }\textbf {\bibinfo {volume} {96}},\ \bibinfo {pages} {012156} (\bibinfo
  {year} {2017})}\BibitemShut {NoStop}%
\bibitem [{\citenamefont {Reimann}\ \emph {et~al.}(2002)\citenamefont
  {Reimann}, \citenamefont {Van~den Broeck}, \citenamefont {Linke},
  \citenamefont {H{\"a}nggi}, \citenamefont {Rubi},\ and\ \citenamefont
  {P{\'e}rez-Madrid}}]{reimann2002}%
  \BibitemOpen
  \bibfield  {author} {\bibinfo {author} {\bibfnamefont {P.}~\bibnamefont
  {Reimann}}, \bibinfo {author} {\bibfnamefont {C.}~\bibnamefont {Van~den
  Broeck}}, \bibinfo {author} {\bibfnamefont {H.}~\bibnamefont {Linke}},
  \bibinfo {author} {\bibfnamefont {P.}~\bibnamefont {H{\"a}nggi}}, \bibinfo
  {author} {\bibfnamefont {J.}~\bibnamefont {Rubi}}, \ and\ \bibinfo {author}
  {\bibfnamefont {A.}~\bibnamefont {P{\'e}rez-Madrid}},\ }\href@noop {}
  {\bibfield  {journal} {\bibinfo  {journal} {Phys. Rev. E}\ }\textbf {\bibinfo
  {volume} {65}},\ \bibinfo {pages} {031104} (\bibinfo {year}
  {2002})}\BibitemShut {NoStop}%
\bibitem [{\citenamefont {Gross}\ \emph {et~al.}(2007)\citenamefont {Gross},
  \citenamefont {Vershinin},\ and\ \citenamefont
  {Shubeita}}]{gross2007CurrBiol}%
  \BibitemOpen
  \bibfield  {author} {\bibinfo {author} {\bibfnamefont {S.~P.}\ \bibnamefont
  {Gross}}, \bibinfo {author} {\bibfnamefont {M.}~\bibnamefont {Vershinin}}, \
  and\ \bibinfo {author} {\bibfnamefont {G.~T.}\ \bibnamefont {Shubeita}},\
  }\href@noop {} {\bibfield  {journal} {\bibinfo  {journal} {Curr. Biol.}\
  }\textbf {\bibinfo {volume} {17}},\ \bibinfo {pages} {R478} (\bibinfo {year}
  {2007})}\BibitemShut {NoStop}%
\bibitem [{\citenamefont {Imparato}(2015)}]{imparato2015}%
  \BibitemOpen
  \bibfield  {author} {\bibinfo {author} {\bibfnamefont {A.}~\bibnamefont
  {Imparato}},\ }\href@noop {} {\bibfield  {journal} {\bibinfo  {journal} {New
  J. Phys.}\ }\textbf {\bibinfo {volume} {17}},\ \bibinfo {pages} {125004}
  (\bibinfo {year} {2015})}\BibitemShut {NoStop}%
\bibitem [{\citenamefont {Reichert}\ and\ \citenamefont
  {Stark}(2005)}]{reichert2005EPJE}%
  \BibitemOpen
  \bibfield  {author} {\bibinfo {author} {\bibfnamefont {M.}~\bibnamefont
  {Reichert}}\ and\ \bibinfo {author} {\bibfnamefont {H.}~\bibnamefont
  {Stark}},\ }\href@noop {} {\bibfield  {journal} {\bibinfo  {journal} {Euro.
  Phys. J. E}\ }\textbf {\bibinfo {volume} {17}},\ \bibinfo {pages} {493}
  (\bibinfo {year} {2005})}\BibitemShut {NoStop}%
\bibitem [{\citenamefont {Golestanian}\ \emph {et~al.}(2011)\citenamefont
  {Golestanian}, \citenamefont {Yeomans},\ and\ \citenamefont
  {Uchida}}]{golestanian2011SoftMatter}%
  \BibitemOpen
  \bibfield  {author} {\bibinfo {author} {\bibfnamefont {R.}~\bibnamefont
  {Golestanian}}, \bibinfo {author} {\bibfnamefont {J.~M.}\ \bibnamefont
  {Yeomans}}, \ and\ \bibinfo {author} {\bibfnamefont {N.}~\bibnamefont
  {Uchida}},\ }\href@noop {} {\bibfield  {journal} {\bibinfo  {journal} {Soft
  Matter}\ }\textbf {\bibinfo {volume} {7}},\ \bibinfo {pages} {3074} (\bibinfo
  {year} {2011})}\BibitemShut {NoStop}%
\bibitem [{\citenamefont {Marchetti}\ \emph {et~al.}(2013)\citenamefont
  {Marchetti}, \citenamefont {Joanny}, \citenamefont {Ramaswamy}, \citenamefont
  {Liverpool}, \citenamefont {Prost}, \citenamefont {Rao},\ and\ \citenamefont
  {Simha}}]{marchetti2013RMP}%
  \BibitemOpen
  \bibfield  {author} {\bibinfo {author} {\bibfnamefont {M.}~\bibnamefont
  {Marchetti}}, \bibinfo {author} {\bibfnamefont {J.}~\bibnamefont {Joanny}},
  \bibinfo {author} {\bibfnamefont {S.}~\bibnamefont {Ramaswamy}}, \bibinfo
  {author} {\bibfnamefont {T.}~\bibnamefont {Liverpool}}, \bibinfo {author}
  {\bibfnamefont {J.}~\bibnamefont {Prost}}, \bibinfo {author} {\bibfnamefont
  {M.}~\bibnamefont {Rao}}, \ and\ \bibinfo {author} {\bibfnamefont {R.~A.}\
  \bibnamefont {Simha}},\ }\href@noop {} {\bibfield  {journal} {\bibinfo
  {journal} {Rev. Mod. Phys.}\ }\textbf {\bibinfo {volume} {85}},\ \bibinfo
  {pages} {1143} (\bibinfo {year} {2013})}\BibitemShut {NoStop}%
\bibitem [{\citenamefont {O’Keeffe}\ \emph {et~al.}(2017)\citenamefont
  {O’Keeffe}, \citenamefont {Hong},\ and\ \citenamefont
  {Strogatz}}]{Strogatz2017NatCommun}%
  \BibitemOpen
  \bibfield  {author} {\bibinfo {author} {\bibfnamefont {K.~P.}\ \bibnamefont
  {O’Keeffe}}, \bibinfo {author} {\bibfnamefont {H.}~\bibnamefont {Hong}}, \
  and\ \bibinfo {author} {\bibfnamefont {S.~H.}\ \bibnamefont {Strogatz}},\
  }\href@noop {} {\bibfield  {journal} {\bibinfo  {journal} {Nat. Commun.}\
  }\textbf {\bibinfo {volume} {8}},\ \bibinfo {pages} {1504} (\bibinfo {year}
  {2017})}\BibitemShut {NoStop}%
\end{thebibliography}%

\clearpage 

\appendix
\setcounter{equation}{0}
\setcounter{figure}{0}
\renewcommand{\theequation}{S\arabic{equation}}
\renewcommand{\thefigure}{S\arabic{figure}}

\section{SUPPLEMENTAL MATERIAL}
{\bf Thermodynamic uncertainty relation for a Stuart-Landau oscillator}
Here, we derive the thermodynamic uncertainty relation between heat dissipation and phase precision for a Stuart-Landau oscillator~\cite{cao2015}:
\begin{equation}
\frac{dz}{dt} = (R^2 + i \omega + i \alpha |z|^2 - |z|^2)z,
\end{equation}
where the complex variable $z = r \exp(i \theta)$ includes amplitude $r$ and phase $\theta$.
The converging amplitude $R$, intrinsic frequency $\omega$, and amplitude-phase coupling $\alpha$ are all positive-definite.
The corresponding amplitude and phase dynamics under noisy environment are
\begin{eqnarray}
\label{eq:amplitude}
\frac{dr}{dt} &=& R^2 r - r^3 + \eta_r \\
\label{eq:phase}
\frac{d \theta}{dt} &=& \omega + \alpha r^2 + \eta_\theta,
\end{eqnarray}
where $\langle \eta_r(t) \rangle = \langle \eta_\theta (t) \rangle =0$, $\langle \eta_r(t) \eta_r(s) \rangle = 2D \delta(t-s)$, and $\langle \eta_\theta(t) \eta_\theta(s) \rangle = (2D/R^2)\delta(t-s)$.
The amplitude of the non-linear oscillator ($r$) converges to $R$.
Then, the phase can be clearly defined without ambiguity only for the small amplitude fluctuation, $\langle \delta r^2 \rangle \equiv \langle (r-R)^2 \rangle \ll R^2$.
Expansion of the above equations around $r = R + \delta r$ gives
\begin{eqnarray}
\frac{d \delta r}{dt} &=& -2R^2 \delta r + \eta_r \\
\frac{d \theta}{dt} &=& \omega + \alpha R^2 + 2\alpha R \delta r + \eta_\theta.
\end{eqnarray}
Then, one can obtain $\lim_{\tau \to \infty} \langle \delta r^2(\tau) \rangle  = D/(2R^2)$, and the condition $\langle \delta r^2 \rangle \ll R^2$ is translated as $2R^4 \gg D$.
Similarly, the phase fluctuation after one period $\tau$ is calculated as
\begin{align}
\langle \delta \theta^2 (\tau) \rangle &= \frac{2D\tau}{R^2}+ (2\alpha R)^2 \int_0^\tau dt_1 \int_0^\tau dt_2  \langle \delta r(t_1) \delta r(t_2) \rangle \nonumber \\
& =\frac{2 D \tau}{R^2}( 1 + \alpha^2 R^2).
\end{align}
Then, the dimensionless phase fluctuation is determined as
\begin{equation}
\label{eq:phase_fluctuation}
\frac{\langle \delta \theta(\tau)^2 \rangle}{\langle \theta(\tau) \rangle^2} = \frac{D \tau}{2\pi^2 R^2}( 1 + \alpha^2 R^2).
\end{equation}

Next, we compute the heat dissipated from the Stuart-Landau oscillator.
The probability density of amplitude and phase $P(r, \theta, t)$ evolves according to the Fokker-Planck equation:
\begin{align}
\frac{\partial P}{\partial t} &= -\frac{1}{r} \frac{\partial}{\partial r} \bigg[ (R^2r - r^3)rP - r D \frac{\partial P}{\partial r} \bigg]  \nonumber \\
&{\phantom{M}} - \frac{\partial}{\partial \theta}\bigg[ (\omega+ \alpha r^2)P - \frac{D}{r^2} \frac{\partial^2 P}{\partial \theta^2} \bigg] \nonumber\\
&= -\frac{1}{r} \frac{\partial J_r}{\partial r} - \frac{\partial J_\theta}{\partial \theta}.
\end{align}
The steady-state probability density $P^{ss}(r, \theta)$ depends only on the amplitude $r$:
\begin{equation}
P^{ss}(r) = Z \exp \bigg[ -\frac{r^4 - 2R^2 r^2}{4D} \bigg]
\end{equation}
with a normalization constant $Z$ to satisfy $2 \pi \int_0^\infty r P^{ss}(r) dr= 1$.
Using $P^{ss}(r)$, one can obtain the probability current for amplitude and phase, $J^{ss}_r=0$ and $J^{ss}_\theta = {\omega_{\text{eff}}}(r) P^{ss}$, respectively, with ${\omega_{\text{eff}}}(r)=\omega+\alpha r^2$.
The mean heat dissipation rate is obtained as
\begin{equation}
\sigma = \int_0^{2\pi} d\theta \int_0^\infty rdr \frac{(rJ^{ss}_\theta)^2}{P^{ss}}= {\langle r^2 {\omega_{\text{eff}}^2}\rangle}.
\end{equation}
Thus, the heat dissipated over a time period $\tau$ is
\begin{align}
q(\tau) &= \tau \sigma= \frac{2 \pi  \langle r^2 \omega_{\text{eff}}^2\rangle}{\langle \omega_{\text{eff}} \rangle} \nonumber \\
&\approx {2\pi R^2 \omega} \bigg( 1+ \frac{\alpha R^2}{\omega} + \frac{2\alpha^2D}{\omega(\omega + \alpha R^2)} + \frac{4\alpha D}{R^2 \omega} \bigg)  \nonumber \\
&\approx {2\pi R^2 \omega} \bigg( 1+ \frac{2\alpha^2D}{\omega^2} + \frac{4\alpha D}{R^2 \omega} \bigg) \nonumber \\
\label{eq:heat_approx}
& \approx {2\pi R^2 \omega},
\end{align}
where we used the condition of small amplitude fluctuation ($2R^4 \gg D$) and another condition for phase dynamics ($\omega \gg \alpha R^2$), which implies that the intrinsic factor $\omega$ should be much greater than the amplitude coupling factor $\alpha R^2$ in $\omega_{\text{eff}} = \omega + \alpha r^2$.
Then, from the two conditions, it follows that $2R^2 \omega \gg \alpha D$ and $2 \omega^2 \gg \alpha^2 D$. 
The last line in Eq.~(\ref{eq:heat_approx}) follows from these two inequalities.
Finally, the cost-precision trade-off can be evaluated using the heat dissipation (Eq.~(\ref{eq:heat_approx})) and the phase fluctuation (Eq.~(\ref{eq:phase_fluctuation})):
\begin{align}
\calQ &\equiv q(\tau) \times \frac{\langle \delta \theta(\tau)^2 \rangle}{\langle \theta(\tau) \rangle^2} \nonumber \\
& =  {2\pi R^2 \omega}  \frac{D 2\pi/\langle \omega_{\text{eff}}\rangle}{2\pi^2 R^2} ( 1 + \alpha^2 R^2) \nonumber \\
&=  2 k_B T (1 + \alpha^2 R^2) \ge 2k_B T, 
\end{align}
where we have used $\langle \omega_{\text{eff}}\rangle /\omega = \langle 1+\alpha r^2/\omega \rangle \approx 1$ under the condition of $\omega \gg \alpha R^2$, and the Einstein relation $D = \mu k_B T$ with $\mu = 1$.
Therefore, the uncertainty relation achieves the minimal bound $\calQ_{\text{min}}=2k_BT$, when the amplitude coupling for phase dynamics is absent ($\alpha=0$).
Although the heat dissipation and phase fluctuation of the noisy Stuart-Landau oscillator have been considered in Ref~\cite{cao2015}, the minimal bound of their trade-off is corroborated in this study for the first time.
The pure phase dynamics corresponds to the noisy Kuramoto model without coupling,
\begin{equation}
\frac{d \theta}{dt} = \omega + \eta,
\end{equation}
is the basis for this study.
\\

{\bf Phase velocity and heat dissipation of coupled oscillators}
Using the probability density $P(\theta_1, \theta_2, t)$ for phases of two coupled oscillators, one can derive the mean phase velocity and mean heat dissipation rate at non-equilibrium steady state~\cite{izumida2016}.
We consider the transformed probability density $P(\theta_1, \theta_2, t) = P(\phi_2, t)/2\pi$ for the orthogonal coordinate, $\phi_1=\theta_1+\theta_2$ and $\phi_2 = \theta_1 - \theta_2$.
The probability density function follows the Fokker-Planck equation:
\begin{equation}
\frac{\partial P(\phi_2,t)}{\partial t}=-\frac{\partial}{\partial \phi_2}\bigg[\Delta \omega -K \sin \phi_2 - 2D \frac{\partial}{\partial \phi_2} \bigg] P(\phi_2, t).
\end{equation}
Its steady state solution is given as 
\begin{equation}
P^{ss}(\phi_2) = \frac{1}{Z} \exp \bigg[ - \frac{V(\phi_2)}{2D} \bigg] \int_{\phi_2}^{\phi_{2}+2\pi} d\psi \exp \bigg[ \frac{V(\psi)}{2D} \bigg],
\end{equation}
where $V(\phi_2) = -K \cos \phi_2 - \Delta \omega \phi_2$ is an effective potential, and $Z$ is a normalization constant to satisfy $\int_0^{2\pi} d\phi_2 P^{ss}(\phi_2) = 1$:
\begin{equation}
Z=\int_0^{2\pi} d\phi_2 \int_{\phi_2}^{\phi_{2}+2\pi} d\psi \exp \bigg[ \frac{V(\psi)-V(\phi_2)}{2D} \bigg].
\end{equation}
Then, using the steady state probability density $P^{ss}$, one can define the steady state current as
\begin{align}
J_i^{ss}(\phi_2) &= \bigg[ F_i  - D \frac{\partial}{\partial \theta_i} \bigg] P^{ss}(\theta_1, \theta_2)\nonumber \\
&= \bigg[ F_i - D\bigg( \frac{\partial \phi_1}{\partial \theta_i} \frac{\partial}{\partial \phi_1}+\frac{\partial \phi_2}{\partial \theta_i} \frac{\partial}{\partial \phi_2} \bigg) \bigg] P^{ss}(\phi_1, \phi_2) \nonumber \\
&= \bigg[ F_i + \frac{1}{2} \frac{\partial \phi_2}{\partial \theta_i} \frac{\partial U(\phi_2)}{\partial \phi_2} \bigg] \frac{P^{ss}(\phi_2)}{2\pi} \nonumber \\
& {\phantom{M}}+ \frac{D}{2\pi Z} \frac{\partial \phi_2}{\partial \theta_i} \bigg[ 1 - \exp\bigg(-\frac{\Delta \omega \pi}{D} \bigg) \bigg].
\end{align}
The steady state current for each oscillator is
\begin{align}
J_1^{ss} = \frac{\omega_1+\omega_2}{4 \pi} P^{ss}(\phi_2)  + \frac{D}{2\pi Z} \bigg[ 1 - \exp\bigg(-\frac{\Delta \omega \pi}{D} \bigg) \bigg], \\
J_2^{ss} = \frac{\omega_1+\omega_2}{4 \pi} P^{ss}(\phi_2)  - \frac{D}{2\pi Z} \bigg[ 1 - \exp\bigg(-\frac{\Delta \omega \pi}{D} \bigg) \bigg].
\end{align}
Given the steady state current, one can obtain the mean phase velocity
\begin{align}
v_i& = \int_0^{2\pi} d\theta_1 \int_0^{2\pi} d\theta_2 J_i^{ss}(\theta_1, \theta_2)\nonumber\\
&= 2\pi \int_0^{2\pi} d\phi_2 J_i^{ss}(\phi_2).
\end{align}
The mean phase velocity for each oscillator is
\begin{align}
v_1&= \frac{\omega_1+\omega_2}{2}  + \frac{2\pi D}{Z} \bigg[ 1 - \exp\bigg(-\frac{\Delta \omega \pi}{D} \bigg) \bigg],\nonumber\\
v_2 &=  \frac{\omega_1+\omega_2}{2}  - \frac{2\pi D}{Z} \bigg[ 1 - \exp\bigg(-\frac{\Delta \omega \pi}{D} \bigg) \bigg].
\end{align}
For $K \to 0$, $v_i \to \omega_i$, whereas for the large $K$ limit, $v_1=v_2=(\omega_1+\omega_2)/2$.
Next, using the steady state current, one can also obtain the heat dissipation rate,
\begin{align}
{\sigma_i} &= \int_0^{2\pi} d\theta_1 \int_0^{2\pi} d\theta_2 F_i(\theta_1, \theta_2) J_i^{ss}(\theta_1, \theta_2) \nonumber \\
&= 2\pi \int_0^{2\pi} d\phi_2  F_i(\phi_2) J_i^{ss}(\phi_2).
\end{align}
The mean heat dissipation rate for each oscillator is
\begin{align}
\sigma_1 &= \omega_1 v_1 -\frac{(\omega_1 + \omega_2)K}{4} \langle \sin \phi_2 \rangle,\nonumber\\
\sigma_2 &=\omega_2 v_2 + \frac{(\omega_1 + \omega_2)K}{4} \langle \sin \phi_2 \rangle,
\end{align}
where $
\langle \sin \phi_2 \rangle \equiv 2\pi \int_0^{2\pi} d\phi_2  \sin \phi_2 P^{ss}(\phi_2)$.
At the large $K$ limit, $\langle \sin \phi_2 \rangle = (\omega_1 - \omega_2)/K$; and hence the two synchronized oscillators have the same heat dissipation rate $\sigma_1=\sigma_2=(\omega_1+\omega_2)^2/4$. On the other hand, in the limit of $K \to 0$, $\sigma_i = \omega_i^2$.
\\

\begin{figure}[h]
\centering
\includegraphics[width=7.cm]{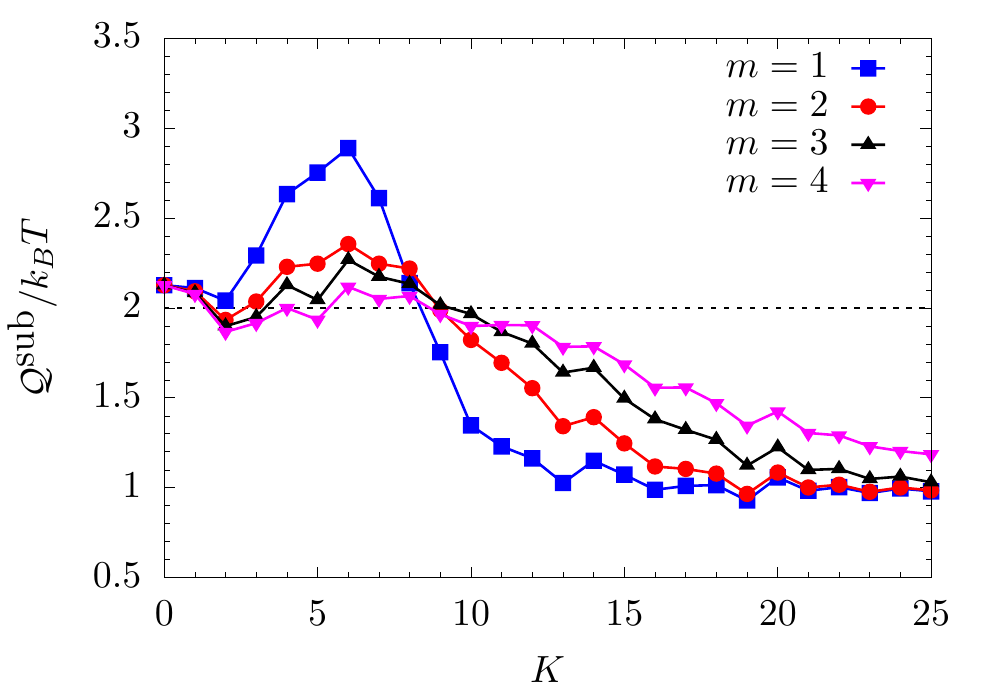}
\caption{Thermodynamic uncertainty relation for higher-order sine interactions, $\sin [m(\theta_j - \theta_i)]$.
${\cal Q}^{\text{sub}}/k_BT$ for various $m \in \{1, 2, 3, 4\}$ for two coupled oscillators ($N=2$) with intrinsic frequencies, $\omega_1=10$ and $\omega_2=5$. 
The noise strength is set to $D=1$.
To compute the cost-precision trade-off, an ensemble of $10^3$ realizations of stochastic process are used.}
\label{figS1}
\end{figure}

{\bf General odd parity interactions}
Our conclusion on the uncertainty relation for interacting oscillators still holds for the higher-order odd trigonometric interactions, $\sin [m(\theta_j - \theta_i)]$ (Fig.~\ref{figS1}).
This implies that the uncertainty relation for interacting oscillators with a periodic coupling with odd parity, of which function is decomposable to a Fourier series in terms of odd harmonics.
\\

\begin{figure}[h]
\centering
\includegraphics[width=8.cm]{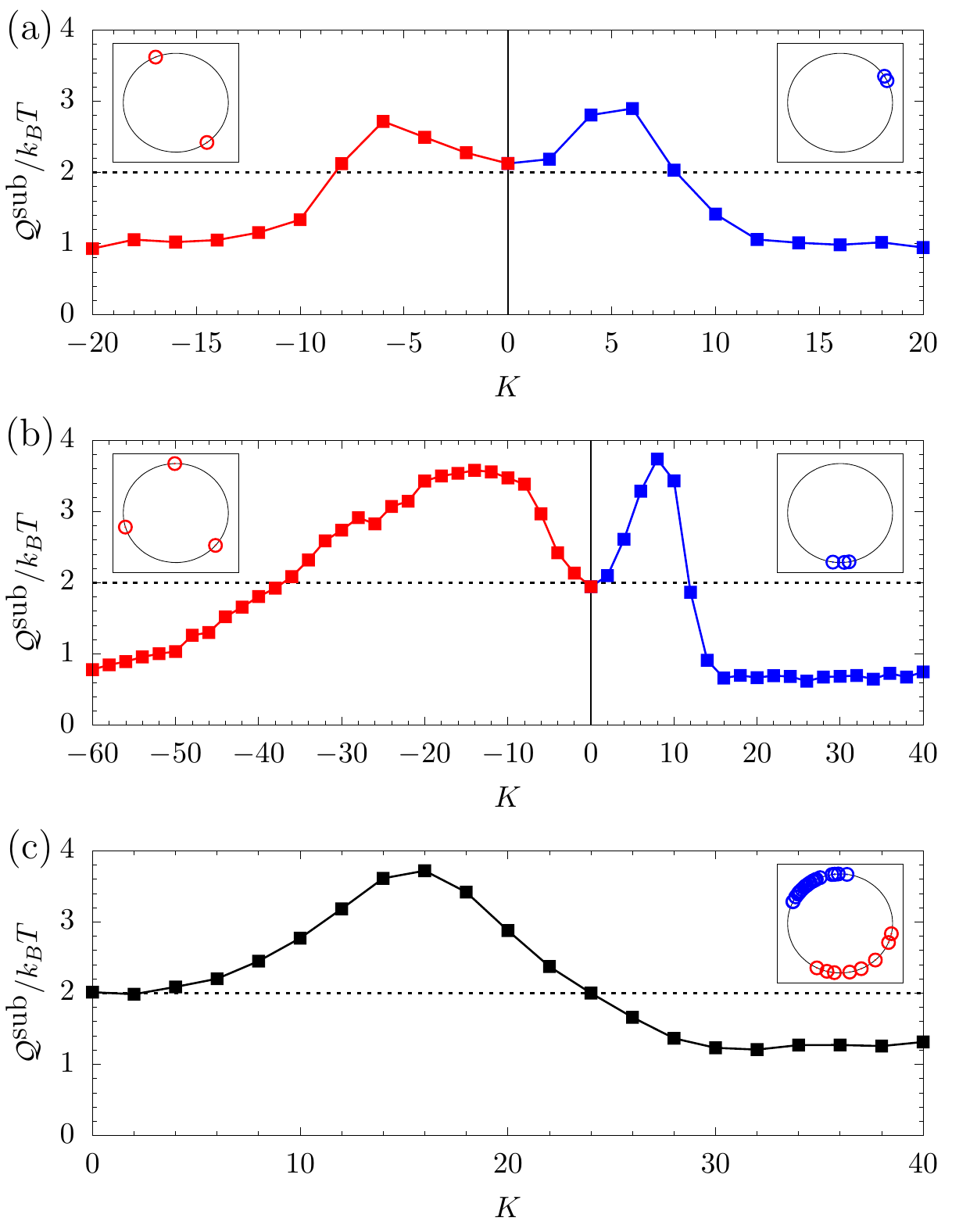}
\caption{Sub-system cost-precision trade-offs for various collective dynamics.
(a) Two coupled oscillators ($N=2$) with intrinsic frequencies $\omega_1=10$ and $\omega_2=5$ under attractive ($K>0$) and repulsive ($K<0$) interactions.
(b) Three coupled oscillators ($N=3$) with intrinsic frequencies $\omega_1=15$, $\omega_2=10$, and $\omega_3=5$ under attractive and repulsive interactions.
(c) Thirty two coupled oscillators ($N=32$) of which intrinsic frequencies are sampled from a normal distribution $\mathcal{N}(\bar{\omega},\Delta \omega)$ with a mean $\bar{\omega} = 10$ and a standard deviation $\Delta \omega=5$. Among them, 16 oscillators (``conformists") have attractive coupling strengths ($K_+ = K$), whereas 8 oscillators (``contrarians'') have negative coupling strengths ($K_-=-K/2$).
The noise strength is set to $D=1$. To compute the trade-off, an ensemble of $10^3$ realizations of stochastic process are used.
Insets represent phase snapshots of attractive oscillators (blue circles) and repulsive oscillators (red circles) at strong coupling regimes.}
\label{figS2}
\end{figure}

{\bf Collective dynamics of coupled oscillators}
To have the uncertainty relation with reduced minimal bound, in-phase synchrony is not a prerequisite.
Even when two repulsive oscillators with negative coupling strengths are synchronized out of phase (Fig.~\ref{figS2}a),
or when three repulsive oscillators repel to each other, forming a ``splay state'' (Fig.~\ref{figS2}b), each oscillator can have a uncertainty bound smaller than $2k_BT$. 
Furthermore, the same conclusion is reached when attractive and repulsive oscillators are mixed to interact to each other to form two clusters, called a ``cluster state'' (Fig.~\ref{figS2}c).

\end{document}